# Pattern Recognition Techniques for the Identification of Activities of Daily Living using Mobile Device Accelerometer

Ivan Miguel Pires[1,2,3], Nuno M. Garcia[1,3,4], Nuno Pombo[1,3,4], Francisco Flórez-Revuelta[5] and Susanna Spinsante[6]

[1]Instituto de Telecomunicações, Universidade da Beira Interior, Covilhã, Portugal
[2]Altranportugal, Lisbon, Portugal
[3]ALLab - Assisted Living Computing and Telecommunications Laboratory, Computer Science Department, Universidade da Beira Interior, Covilhã, Portugal
[4]Universidade Lusófona de Humanidades e Tecnologias, Lisbon, Portugal
[5]Department of Computer Technology, Universidad de Alicante, Spain
[6]Università Politecnica delle Marche, Ancona, Italy

impires@it.ubi.pt, ngarcia@di.ubi.pt, ngpombo@di.ubi.pt, francisco.florez@ua.es, s.spinsante@univpm.it

**Abstract**

This paper focuses on the recognition of Activities of Daily Living (ADL) applying pattern recognition techniques to the data acquired by the accelerometer available in the mobile devices. The recognition of ADL is composed by several stages, including data acquisition, data processing, and artificial intelligence methods. The artificial intelligence methods used are related to pattern recognition, and this study focuses on the use of Artificial Neural Networks (ANN). The data processing includes data cleaning, and the feature extraction techniques to define the inputs for the ANN. Due to the low processing power and memory of the mobile devices, they should be mainly used to acquire the data, applying an ANN previously trained for the identification of the ADL. The main purpose of this paper is to present a new method implemented with ANN for the identification of a defined set of ADL with a reliable accuracy. This paper also presents a comparison of different types of ANN in order to choose the type for the implementation of the final method. Results of this research probes that the best accuracies are achieved with Deep Learning techniques with an accuracy higher than 80%.

## 1. Introduction

Accelerometer is a sensor commonly available in off-the-shelf mobile devices [1] that measures the acceleration of the movement of the mobile device, allowing the creation of a method for the recognition of Activities of Daily Living (ADL) [2]. After the development of a method for the identification of ADL, it could be integrated in the creation of a personal digital life coach [3], important for the monitoring of elderly people, and people with some type of impairment, or for the training the people's lifestyle.

The methods related to the recognition of the ADL with accelerometer may be used for the recognition of the daily activities with movement, including running, walking, walking on stairs, and standing. Following the previous research studies [4-6], the recognition of the ADL is composed by several steps, such as the data acquisition, the data processing, composed by data cleaning, data imputation, and feature extraction, the data fusion, and the artificial intelligence methods for the concrete identification of the ADL. However, this study only uses the accelerometer sensor, removing some steps of the proposed architecture. Based on the assumption that the sensor was always acquiring the data, the final steps used are the data acquisition, the data cleaning, and the application of the artificial intelligence methods.

During the last years, the recognition of ADL has been studies by several authors [7-12], verifying that the Artificial Neural Networks (ANN) are widely used. This paper proposes the creation of a method for the recognition of ADL using accelerometer, comparing three types of ANN, such as Multilayer Perception (MLP) with Backpropagation, Feedforward neural network with Backpropagation, and Deep Learning, in order to verify the method that achieves the best accuracy in the recognition of running, walking, going upstairs, going downstairs, and standing. The datasets are composed with raw accelerometer data acquired by individual with ages ranged between 16 and 60 years old and different lifestyles, with a mobile device in the pocket. For the implementation of these types of ANN are used several datasets with different sets of features, identifying the best features to increase the accuracy of the recognition, and three Java libraries are used for the implementation of the different methods, such as Neuroph [13], Encog [14], and DeepLearning4j [15], achieving the best accuracy with Deep Learning methods.

The remaining sections of this paper are organized as follows: Section 2 presents a brief literature review related to the identification of ADL using accelerometer. Section 3 presents the methodology used for the creation of a solution for the recognition of the ADL using the accelerometer sensor. Section 4 presents the results obtained during the research presented. In section 5, the discussion and conclusions about the results are presented.

## 2. Related Work

The identification of the Activities of Daily Living (ADL) [2] may be performed with several methodologies, including the use of the raw data as input or the extraction of several features that can be used as input for the implemented method.

The authors of [16] used the accelerometer embedded on a mobile device to develop a method for the recognition of the resting and walking states, which implements an Artificial Neural Network (ANN) that receives as input the raw data acquired from the accelerometer, reporting an accuracy around 95%.

However the use of features extracted from the accelerometer data is most common. In [17], the authors extracted some features from the accelerometer signal, these are: mean along z-axis, maximum, minimum, standard deviation and Root Mean Square (RMS) from the magnitude of the acceleration, average of peak frequency (APF), standard deviation, RMS, maximum and minimum along x-axis, y-axis and z-axis, and correlation between z-axis and y-axis. Then, the authors applied several methods to test the accuracy for the identification of running, slow walking, fast walking, aerobic dancing, going up stairs, and going down stairs, these are: ANN (i.e., Multilayer Perceptron (MLP)), Support Vector Machine (SVM), Random Forest, Logistic Model Trees (LMT), and Simple Logistic Logit Boost. The authors reported an accuracy of 89.72%, applying the MLP method [17].

The authors of [18] developed a method to identify some ADL, such as walking, running, going up stairs, going down stairs, driving, cycling and standing, using accelerometer data. The method extracted several features from the magnitude of the acceleration, these are: mean, standard deviation, cross-axis signals correlation, Fast Fourier Transform (FFT) spectral energy, frequency domain entropy and log of FFT, and implements the Naïve Bayes, C4.5 Decision Tree, K-Nearest Neighbor (KNN), and SVM methods, reporting an average of true positives greatest than 95% [18].

The accelerometer sensor is also used in [19] for the recognition of going up stairs, going up on an escalator, and walking on a ramp, implementing decision tables, J48 Decision tree, Naïve Bayes, KNN and MLP methods with several features, including the mean, standard deviation, skewness, kurtosis, average absolute deviation, and pairwise correlation of the tree axis, and the average of the resultant acceleration, achieving 94% of recognition accuracy.

In [20], the authors implemented a method using decision tree for the recognition of standing, walking, running and going up stairs and going down stairs, with the mean, median, variance, standard deviation, maximum, minimum, range, RMS, FFT coefficients, and FTT spectral energy as features, reporting an accuracy between 73.72% to 88.32%. The authors of [21] also implemented a decision tree method for the recognition of walking, running, sitting, and standing, based on the mean, variance, bin distribution in time and frequency domain, FFT spectral energy, and correlation of the magnitude of the accelerometer data, reporting an accuracy of 98.69%.

The authors of [22] implemented several methods, such as Bayesian Network, MLP, Naïve Bayes, C4.5 decision tree, Random Tree, Radial Basis Function Network, Sequential Minimal Optimization (SMO), and Logistic Regression, using the accelerometer data for the recognition of walking, jogging, going up stairs, going down stairs, sitting, standing, and laying down activities, with the mean, standard deviation, mean absolute deviation, time between peaks and resultant magnitude, of the magnitude of the acceleration as features. With Bayesian Network, the authors reported a maximum accuracy of 77.81%; with MLP, the maximum accuracy reported is 94.44%; with Naïve Bayes, the maximum accuracy reported is 58.06%; with C4.5 decision tree, the maximum accuracy reported is 95.40%; with Random Tree, the maximum accuracy reported is 94.67%; with Radial Basis Function Network, the maximum accuracy reported is

73.03%; with Sequential Minimal Optimization (SMO), the maximum accuracy reported is 90.27%; and with Logistic Regression, the maximum accuracy reported is 92.71%.

The walking, running, standing and sitting are also recognized by [9], using the accelerometer data and implementing the Clustered KNN method, with the mean, minimum, maximum and standard deviation of the magnitude of the acceleration as features, reporting an overall accuracy around 92%. The same activities are also recognized by [23], using mean, maximum, minimum, median, and standard deviation of the magnitude of acceleration as features, implementing the MLP method, and reporting an accuracy greatest than 99% for any activity.

In [24], a system that implements the Hidden Markov Model Ensemble (HMME) method for the accelerometer data in order to recognize walking, going up stairs, going down stairs, sitting, standing and laying activities, using the mean and standard deviation as features, reporting an accuracy of 83.55%. The authors of [25] also implemented the Hidden Markov Model (HMM), decision tree and Random Forest methods with accelerometer data for the identification of walking, going up stairs, and going down stairs. The features used are mean, variance, standard deviation, median, minimum, maximum, range, Interquartile range, Kurtosis, skewness and spectrum peak position of the accelerometer data, reporting an accuracy of 93.8% [25]. Another authors implemented the Sliding-Window-based Hidden Markov Model (SW-HMM), and compared this method with SVM and ANN for the recognition of walking, standing, running, going up stairs, and going down stairs activities, using the mean, variance and quartiles of the accelerometer data [26], reporting an accuracy around 80%.

The J48 decision tree, Random Forest, Instance-based learning (IBk), and rule induction (J-Rip) methods were used with accelerometer data for the recognition of standing, sitting, going up stairs, going down stairs, walking, and jogging, implementing the Dual-tree complex wavelet transform (DT-CWT), DT-CWT statistical information and orientation as features, reporting an accuracy of 86% for the recognition of all activities [27].

The authors of [28] created a system named Actitracker, that performs the recognition of walking, jogging, going up stairs, going down stairs, standing, sitting, and lying down activities, using the Random Forest method and accelerometer data. This systems uses the mean and standard deviation for each axis, the bin distribution and the heuristic measure of wave periodicity, with an accuracy around 90% [28].

In [10], the authors implemented a solution using ANN and SVM methods applied to the accelerometer data, in order to identify several activities, such as standing, sitting, standing up from a chair, sitting down on a chair, walking, lying, and falling activities. The features implemented are sum of all magnitude of the vectors, sum of all magnitude of the vectors excluding the gravity, maximum and minimum value of acceleration in gravity vector direction, mean of absolute deviation of acceleration in gravity vector direction, and gravity vector changing angle, reporting that the results have a sensitivity of 96.67% and specificity of 95% [10].

In order to the recognition of going up stairs, going down stairs, walking, jogging, and jumping activities, the authors of [29] used the KNN, Random Forests and SVM methods with accelerometer data to identify accurately the activities. The features extracted are mean, standard deviation, maximum, minimum, correlation, interquartile range, Dynamic time warping distance (DTW), FFT coefficients and wavelet energy, reporting an accuracy around 95% [29].

In [30], the authors proposed a solution that uses SVM, J48 decision tree and Random Forest methods with accelerometer data for the recognition of Sitting, Standing, Walking, and running. The solution extracts several features, such as average of peak values, average of peak rising time, average of peak fall time, average time per sample, and average time between peaks, reporting an accuracy of 98.8283% [30].

Other features are used by other authors in [31], including the mean, the median, the standard deviation, the skewness, the kurtosis, the minimum, the maximum, and the slope for each axis and for the absolute value of accelerometer. These features are used to recognize standing, walking, jogging, jumping, going up stairs, going down stairs, and sitting on a chair with the accelerometer data, implementing IBk, J48 decision tree, Logistic regression, and MLP methods, reporting an accuracy of 94% with the implementation of the IBk classifier [31].

The authors of [32] developed a method to recognize a walking pattern, where the person start walking, stop at the red light, going across the road at slightly higher speed on green light, and continuing straight forward or turning left/right, using ANN only with the mean and standard deviation of the accelerometer signal, reporting an accuracy between 75% and 98%.

In [33], the accelerometer data was used to recognize walking, sitting, standing, going up stairs, and going down stairs, after the extraction of the mean, standard deviation, variance, FFT energy, and FFT information entropy from the accelerometer data, applying the decision tree, KNN, and SMO methods.

With decision tree, the maximum accuracy reported by the authors is 91.37%; with KNN, the maximum accuracy reported is 94.29%; and with SMO, the maximum accuracy reported is 84.42%.

In [34], the authors implemented the decision tree, the Bayesian Network, the Naïve Bayes, the KNN, and the rule based learner methods for the recognition of Walking, Standing, Sitting, going up stairs and going down stairs, based on the mean, standard deviation, and correlation of the accelerometer raw data, the energy of FFT, and the mean and standard deviation of the FFT components in the frequency domain. With the Bayesian Network, the maximum accuracy reported by the authors is 95.62%; with the Naïve Bayes, the maximum accuracy reported is 97.81%; with the KNN, the maximum accuracy reported is 99.27%; and with the rule based learner, the maximum accuracy reported is 93.53%.

Another study [35] presents a solution to recognize walking, jogging, cycling, going up stairs, and going down stairs, implementing a decision tree and a probabilistic neural network (PNN) with some features, such as average of the acceleration, standard deviation for each axis, binned distribution for each axis, and average energy for each axis, reporting results with an average accuracy of 98% with the use of accelerometer.

The authors of [36] extracted some features, such as mean, standard deviation, and variance of the accelerometer signal, and implemented the KNN, decision tree, rule-based and MLP methods to recognize walking, sitting, standing, going up stairs and going down stairs activities, verifying that MLP has an accuracy up to 80%.

In [37], the authors used the mean, standard deviation, correlation, mean absolute value, standard deviation absolute value, and power spectral density of the accelerometer data, in the Naïve Bayes, KNN, Decision Tree, and SVM methods for the recognition of walking, cycling, running, and standing activities, reporting an accuracy higher than 95%.

The authors of [38] implement a method based on the peak values of the accelerometer signal, extraction some features, including the number of peaks every 2 seconds, the number of troughs every 2 seconds, the difference between the maximum peak and the minimum trough every 2 seconds, and the sum of all peaks and troughs, in order to recognize walking, jogging, and marching activities. They implemented the J48 decision tree, bagging, decision table, and Naïve Bayes methods, reporting an accuracy of 94% [38].

In [39], the authors implemented a decision tree classifier with several features, such as mean, median, maximum, minimum, RMS, standard deviation, median deviation, interquartile range, energy, entropy, skewness, and kurtosis of the accelerometer data, for the recognition of running, walking, standing, sitting, and laying activities with a reported accuracy of 99.5%.

In [40], the authors implemented a SVM method with several features, such as RMS, variance, correlation and energy of the accelerometer data, for the recognition of walking, running, cycling, and hopping with a reported average accuracy of 97.69%.

The authors of [41] implemented an ANN with mean, standard deviation, and percentiles of the magnitude of the accelerometer data as features, with a reported accuracy of 92% in the recognition of standing, walking, running, going up stairs, going down stairs, and running.

In addition, the authors of [42] implemented an ANN with some features, such as mean, maximum, minimum, difference between maximum and minimum, standard deviation, RMS, Parseval's Energy, correlation between axis, kurtosis, skewness, ratio of the maximum and minimum values in the FFT, difference between the maximum and minimum values in the FFT, median of peaks, median of troughs, number of peaks, number of troughs, average distance between two consecutive peaks, average distance between two consecutive troughs, and ratio of the average values of peaks and troughs based on a window of the accelerometer data. The activities recognized by the method are resting, walking, cycling, jogging, running, and driving [42], reporting an accuracy between 57.53% to 97.58%.

The authors of [43] implemented a SVN method with mean, minimum, maximum, standard deviation, energy, mean absolute deviation, binned distribution, and percentiles of the magnitude of acceleration as features, with a reported overall accuracy of 94.3% in the recognition of running, staying, walking, going up stairs, and going down stairs.

In [44], a method that combines the J48 decision tree, MLP, and Likelihood Ratio (LR) models was implemented and it uses the accelerometer data for the extraction of the minimum, maximum, mean, standard deviation, and zero crossing rate for each axis, and the correlation between axis for the application in the model, in order to recognize the going down stairs, jogging, sitting, standing, going up stairs, and walking activities with a reported accuracy of 97%.

The authors of [45] used the most common features, such as mean, variance, covariance, energy, and entropy of the magnitude of the accelerometer data, and implemented an ANN for the identification of the activity related to play foosball with a reported accuracy higher than 95%.

Another authors [46] also developed a method to identify the activity related to shooting with the implementation of Random Forest, SVM and KNN methods, using the mean, standard deviation, median, maximum, minimum, zero crossing rate, number of peaks, correlation coefficient, and FFT coefficients as features, obtaining a reported accuracy of 91.34%.

In [47], the authors used the accelerometer data and extracted the mean, maximum, minimum, median, standard deviation, Signal Magnitude Area (SMA), mean deviation, Principal Component Analysis (PCA), Interquartile range, skewness and kurtosis of the magnitude of the accelerometer data, implementing the SVM, MLP, Naïve Bayes, KNN, tree, and kStart methods to recognize sitting, standing, laying, walking, and jogging activities, obtaining a reported accuracy of 99.01% with KNN and kStar methods.

The SVM method was applied to the accelerometer data by the authors of [48] in order to recognize the running, walking, sitting, going up stairs, and going down stairs activities, using the mean, variance, standard deviation, median, maximum, minimum, RMS, zero crossing rate, skewness, kurtosis, entropy, and spectral entropy of the magnitude of the accelerometer data, and the correlation between two axis, reporting an accuracy between 60% to 100%, depending on the orientation and placement of the smartphone.

The authors of [49] created a method that uses deep neural networks (DNN) for the identification of walking, running, standing, sitting, lying, going up stairs, and going down stairs activities, using the mean, minimum, maximum, and standard deviation for the three axis and the magnitude of the accelerometer data, reporting an accuracy between 77% and 99%.

In [50], the authors extracted the mean, median, maximum, minimum, RMS, standard deviation, median deviation, interquartile range, minimum average, maximum average, maximum peak height, average peak height, mean cross count, entropy, FFT spectral energy, skewness, and kurtosis from the accelerometer data as features, applying the J48 decision tree, SMO, and Naïve Bayes classifiers in order to recognize the walking, standing, travel by car, travel by bus, travel by train, and travel by metro, reporting an accuracy of 95.6% with J48 decision tree, and 92.4% with SMO in the recognition of all activities.

The authors of [51] extracted only the mean, variance and correlation for all axis of the accelerometer data, applying the Naïve Bayes, MLP, J48 decision tree, and SVM for the recognition of all activities when the user is playing tennis, reporting an accuracy of 100% with SVM classifier.

In order to recognize the walking, jogging, going up stairs, going down stairs, sitting and standing activities, the authors of [52] used the accelerometer data to extract the mean, standard deviation, average absolute difference, average resultant acceleration, time between peaks and binned distribution, and applying the J48 decision tree, logistic regression, MLP, and Straw Man classifiers, reported that the best accuracy was achieved with MLP equal as 97.7%.

In [53], the authors implemented the discrete wavelet transform (DWT) to extract the energy and variances of the coefficients for the recognition of the walking, jogging, going up stairs, going down stairs, sitting, and standing activities, recurring to the SVM, KNN, Bayesian Network, decision tree, and ANN methods, reporting an overall accuracy between 86.3% and 88.8%.

The studies previously presented use only the accelerometer sensor, however another studies used more than the accelerometer, but the features extracted can be separated.

In [11], the authors used the accelerometer and the gyroscope to recognize cleaning, cooking, medication, sweeping, washing hands, and watering plants, implementing MLP, Naïve Bayes, Bayesian Network, Decision table, Best-First Tree, and kStar, but the features related to the accelerometer are mean, maximum, minimum, standard deviation, zero crossing rate, and correlation, and the reported overall accuracy of the method is higher than 90%.

Another authors, in [54], also used the accelerometer and gyroscope sensors to recognize standing, walking, running, going up stairs, going down stairs, and laying, implementing the J48 decision tree, logistic regression, MLP, and SVM methods, and the features related to the accelerometer are the mean, energy, standard deviation, correlation, and entropy, reporting an accuracy between 89.3% and 100%.

The authors of [12] implemented the Random Forest, SVM, MLP, and KNN methods with the accelerometer and gyroscope data for the recognition of going up stairs, going down stairs, walking, running, and jumping, and the features related to the accelerometer sensor are the mean, standard deviation, maximum, minimum, correlation, interquartile range, DTW distance, FFT coefficients, and wavelet energy, reporting an accuracy between 84.97% and 90.65%.

In [55], the authors implemented the SVM method with some features of the accelerometer and gyroscope data, where the features related to the accelerometer are mean, variance, maximum, and minimum along the Y axis, in order to recognize the walking, running, going up stairs, going down stairs, standing, sitting, and cycling, reporting an accuracy of 96%.

The authors of [56] used the accelerometer and the gyroscope for the recognition of falling activities, implementing the Naïve Bayes, Least Squares Method (LSM), MLP, and SVM classifiers with several features related to the accelerometer, such as mean, skewness, maximum, minimum, kurtosis, standard deviation, interquartile range, and median, reporting an accuracy of 87.5%.

The mean and standard deviation of the accelerometer data and the gravity are extracted by authors of [57] for the identification of walking, going up stairs, going down stairs, standing, and lying, with the use of several methods, including Random Forest, SVM, Naïve Bayes, J48 decision tree, MLP, and KNN, reporting an accuracy higher than 90%.

The recognition of walking, sitting, standing, jogging, cycling, going up stairs, and going down stairs was performed by the authors of [58], using accelerometer and gyroscope features, such as standard deviation, minimum, maximum, median, percentiles, sum of the values above or below percentiles, square sum of values above or below percentiles, number of crossings above or below percentiles, and sum of sequences of FFT signals, reporting a reliable accuracy with Linear discriminant analysis (LDA) and Quadratic discriminant analysis (QDA), reporting an accuracy between 83.9% and 96.2%.

In [59], the authors used the accelerometer and the gyroscope sensors, but extracted several features related to the accelerometer, such as mean, and standard deviation for each axis and acceleration magnitude, sum of acceleration magnitude, and FFT magnitude, in order to recognize walking, jogging, sitting, going up stairs, and going down stairs with J48 decision tree, MLP, Naïve Bayes, Logistic regression, and KNN methods, reporting an accuracy between 90.1% and 94.1%.

The authors of [60] implemented the ANOVA method with RMS, maximum, minimum, and zero crossing rates for each axis of the accelerometer in order to identify the resting, sitting, walking, and turning, reporting an accuracy around 100%.

The Naïve Bayes and decision tree classifiers are implemented by the authors of [61] in order to recognize walking, going up stairs, going down stairs, sitting, standing, and laying, with the accelerometer and gyroscope data, where the features related to the accelerometer are mean, standard deviation, median absolute deviation, maximum, minimum, energy of segment, interquartile range, entropy, auto regression coefficient, correlation coefficients, index of the frequency component with largest magnitude, skewness, kurtosis, energy and angle of the magnitude of the acceleration, reporting an accuracy between 79.4% and 96.2%.

The authors of [62] also used the accelerometer and gyroscope data with SVM, KNN, Kernel-Extreme Learning Machine, and Sparse Representation Classifier for the recognition of walking, going up stairs, going down stairs, running, and standing, using several features, such as absolute mean, median absolute deviation, maximum, minimum, Signal Magnitude Range, interquartile range, Kurtosis, and skewness, reporting an average accuracy of 94.5%.

The SVM method was implemented in [63] with variance, mean, maximum, and minimum of the three axis of the accelerometer, reporting an average accuracy of 95.5% in the recognition of walking and jogging activities. The SVM method is also implemented in [64] with the mean, variance, energy, and entropy of the accelerometer data for the recognition of walking, going up stairs, going down stairs, standing, sitting, and laying.

For the identification of cycling, running, sitting, standing, and walking, the authors of [65] tested several classifiers, such as KNN, parse approximation, SVM, Fuzzy c-means, Spearman correlation, linear regression, and MLP, with the mean, standard deviation, median, peak acceleration, power ratio of the frequency bands, and energy of the accelerometer data, reporting an accuracy of 98%.

The SVM and KNN methods was implemented by the authors of [66], using only the mean and standard deviation as accelerometer features for the recognition of resting, walking, going up stairs, going down stairs, and running, reporting an accuracy higher than 90%.

The SVM and Random Forest methods were implemented by the authors of [67], using several features, including the angle of the accelerometer data, the energy, the minimum acceleration value, the maximum acceleration value, and the mean of the acceleration value for the recognition of walking, going up stairs, going down stairs, sitting, standing, and laying, reporting an accuracy around 100% with the use of the accelerometer and the gyroscope sensors.

The SVM method is also implemented with accelerometer and Global Positioning System (GPS) data for the recognition of walking, standing, and running activities, but the accelerometer features used are minimum, maximum, mean, standard deviation, correlation, and median crossing [68], reporting an accuracy of 97.51%.

Another study that uses SVM method makes use of accelerometer, gyroscope, and barometer sensors for the identification of walking, going up stairs, going down stairs, standing, going elevator up, and going elevator down, extraction the mean, mean of 1$^{st}$ half, mean of 2$^{nd}$ half, difference of means, slope, variance, standard deviation, RMS, and Signal Magnitude Area [69], reporting an accuracy between 87.45% and 99.25%.

The SVM method is also used with the accelerometer, the gyroscope, the barometer, and the GPS sensors for the identification of sitting, standing, washing dishes, walking, going up stairs, going down stairs, cycling, and running, using the mean, standard deviation, mean squared, and interquartile range of the accelerometer data as features [70], reporting an accuracy around 90%.

The decision tree classifier is used with accelerometer, GPS, and camera data for the identification of sitting, standing, lying, riding an elevator, walking, dinning, going up stairs, going down stairs, moving a kettle, washing dishes, toast bread, preparing a meal, drying hands, moving dishes, washing hands, brushing teeth, and combing hair, with some acceleration features, such as acceleration along Y axis, standard deviation for each axis, range of Y axis, SMA of the sum of ranges, difference of range, and range of X axis and Z axis, reporting an accuracy between 73% and 100% [71].

The decision tree is also used with accelerometer, gyroscope, magnetometer, and barometer for the identification of going up stairs, going down stairs, taking the elevator, standing, and walking on an escalator, with some features, such as mean, mean of absolute values, median, variance, standard deviation, 75$^{th}$ percentile, interquartile range, average absolute difference, binned distribution, energy, SMA, zero-crossing rate, number of peaks, absolute value of short-time FFT, and entropy, reporting an accuracy performance between 80% and 90% [72].

On the other hand, the authors of [73] implemented decision tree and SVM methods using the microphone and accelerometer data to recognize sitting, lying, standing, walking, going up stairs, going down stairs, jogging, and running, with some features related to the accelerometer data, such as minimum, differences, mean, standard deviation, variance, correlation, coefficient sum, spectral energy, and spectral entropy, reporting an achieved accuracy of 89.12%.

Related to the use of the accelerometer and gyroscope data, the authors of [74] used the Threshold Based Algorithm (TBA) with the values of the acceleration, and the difference between adjacent elements of the heading as features, recognizing the activities related to walking, going up stairs, going down stairs, jumping, and running, reporting an accuracy of 83%.

The Random Forest classifier is also used with the magnetometer, gyroscope, and accelerometer for the recognition of sitting, standing, walking, going up stairs, going down stairs, and running, based on the mean, variance, frequency of the point with maximum amplitude, value of the point with maximum amplitude, energy, and mean of the extremum value [75], reporting an average accuracy of 95.7%.

The authors of [76] extracted the mean, standard deviation, median, skewness, kurtosis, and interquartile range from the accelerometer available on a system composed by light, proximity, accelerometer, and gyroscope sensors, for the identification of standing, walking, running, going up stairs, and going down stairs, using J48 decision tree, Naïve Bayes, and SMO methods, reporting an accuracy of 89.6% using the J48 decision tree classifier.

In [77], the accelerometer and GPS sensor were used with several features, including the mean, FFT energy, and variance of the accelerometer data, to recognize the standing, travelling by car, travelling by train, and walking activities, with several methods, such as J48 decision tree, Random Forest, ZeroR, logistic regression, decision table, Radial basis function (RBF) network, MLP, Naïve Bayes, and Bayesian network, reducing the battery consumption, reporting an accuracy of 95.7%.

The authors of [78] only extracted the mean and standard deviation as features related to the accelerometer data, but the system uses the GPS, accelerometer, magnetometer, and gyroscope, implementing Naïve Bayes, SVM, MLP, Logistic regression, KNN, rule based, and decision tree classifiers to identify the walking, running, sitting, standing, going up stairs, and going down stairs, reporting an accuracy between 69% and 99%.

In [79], the authors used the accelerometer, gyroscope, and camera for the identification of several activities, such as lying, walking, running, jogging, cycling, vacuum cleaning, washing dishes, brushing teeth, going up stairs, going down stairs, eating, playing cello, playing piano, playing tennis, working on a computer, reading a book, folding laundry, doing laundry, wiping cupboard, driving a car, and taking an elevator, using the Naïve Bayes, SVM and HMM methods. For these methods was extracted some features, such as mean and variance for each accelerometer axis, movement intensity, and energy, reporting an accuracy of 81.5% [79].

The authors of [80] used the accelerometer and the microphone to recognize some activities, such as shopping, waiting in a queue at a cash desk of a supermarket, driving, travelling by car, vacuum cleaning, cooking, washing dishes, working on a computer, sleeping, watching TV, and being in a restaurant. These authors extracted the sum of all acceleration values, mean, standard deviation, maximum, minimum, range, coefficient of variation and angular degree of the accelerometer data, applying them to the J48 decision tree, LMT, and IBk classifiers, reporting that the best accuracy was achieved by J48 decision tree with 92.1278% [80].

In [81], the authors developed a system with accelerometer and barometer data, which implements a Least squares support vector machine (LS-SVM)-based on sequential forward selection (SFS) method for the identification of standing, walking, running, going up stairs, going down stairs, and taking an elevator. The features extracted for the implementation of the method is the variance of the total acceleration, RMS of the horizontal acceleration, skewness of the vertical acceleration, and range of the vertical acceleration [81], reporting an accuracy up to 90.7%.

Following the studies analysed, the table 1 shows the distribution of the different ADL recognized, verifying that the walking, resting/standing, going up stairs, going down stairs, running, and sitting are the most recognized ADL.

*Table 1 - Distribution of the ADL extracted in the studies analyzed*

| ADL: | Number of Studies: |
|---|---|
| walking | 63 |
| resting/standing | 48 |
| going up stairs | 45 |
| going down stairs | 44 |
| running | 31 |
| sitting | 30 |
| jogging | 16 |
| laying down | 15 |
| cycling | 10 |
| driving | 6 |
| taking an elevator | 5 |
| jumping | 4 |
| cleaning | 3 |
| washing dishes | 3 |
| going up on an escalator | 2 |
| falling | 2 |
| cooking | 2 |
| dancing | 1 |
| hopping | 1 |
| medication | 1 |
| sweeping | 1 |
| washing hands | 1 |
| watering plants | 1 |
| dinning | 1 |
| brushing teeth | 1 |
| eating | 1 |
| shopping | 1 |
| sleeping | 1 |

Regarding the ADL recognized in the studies analyzed, the table 2 shows the distribution of the different features used, verifying that the mean, minimum, maximum, standard deviation, correlation, median, FFT spectral energy, and variance are the most used features, with more relevance for mean and standard deviation.

*Table 2 - Distribution of the features extracted in the studies analyzed*

| Features: | Number of Studies: |
|---|---|
| Mean (Z axis, X axis, Y axis, Acceleration, Velocity, Gravity, Peaks, Toughs) | 59 |
| Standard Deviation (Acceleration, X axis, Y axis, Z axis, Gravity) | 49 |
| Minimum (Acceleration, X axis, Y axis, Z axis) | 29 |
| Maximum (Acceleration, X axis, Y axis, Z axis) | 29 |
| FFT spectral energy (Acceleration) | 25 |
| Correlation (X axis, Y axis, Z axis) | 19 |
| Variance (Acceleration, X axis, Y axis, Z axis) | 19 |
| Median (Acceleration, Peaks, Toughs) | 18 |
| Skewness (Acceleration, X axis, Y axis, Z axis) | 13 |
| Kurtosis (Acceleration, X axis, Y axis, Z axis) | 12 |
| Inter-quartile-Range (Acceleration) | 12 |
| Root Mean Square (Acceleration, X axis, Y axis, Z axis) | 11 |
| Entropy (Acceleration) | 11 |
| Zero-Cross (Acceleration) | 6 |
| Average Absolute Deviation (X axis, Y axis, Z axis) | 6 |
| bin distribution in time and frequency domain (Acceleration) | 6 |
| Range (Acceleration, X axis, Y axis, Z axis) | 5 |
| Fourier transform coefficients (Acceleration) | 5 |
| Percentiles (10, 25, 75, and 90) (Acceleration) | 4 |
| Sum (Acceleration, Peaks, Troughs) | 4 |
| Count of peaks (Acceleration) | 4 |
| Signal Magnitude Area (SMA) (Acceleration) | 4 |
| Slope (Acceleration) | 3 |
| Dynamic time warping distance (Acceleration) | 3 |
| Difference of means (Acceleration) | 2 |
| time between peaks (Acceleration) | 2 |
| Average Peak value (Acceleration) | 2 |
| Count of troughs (Acceleration) | 2 |
| Mean of 1st half (Acceleration) | 1 |
| Mean of 2st half (Acceleration) | 1 |
| Log of FFT (Acceleration) | 1 |
| Covariance (Acceleration, X axis, Y axis, Z axis) | 1 |
| Spectrum peak position (Acceleration) | 1 |
| Quartiles (Acceleration) | 1 |
| Sum of values above or below percentile (10, 25, 75, and 90) (Acceleration) | 1 |
| Square sum of values above or below percentile (10, 25, 75, and 90) (Acceleration) | 1 |
| Number of crossings above or below percentile (10, 25, 75, and 90) (Acceleration) | 1 |
| Sums of smaller sequences of Fourier-transformed signals (Acceleration) | 1 |
| Gravity vector changing angle (Acceleration) | 1 |
| Average Peak rising time (Acceleration) | 1 |
| Average Peak fall time (Acceleration) | 1 |
| Average Time per sample (Acceleration) | 1 |
| Average Time between peaks (Acceleration) | 1 |
| Difference between the maximum peak and minimum trough (Acceleration) | 1 |
| Principal Component Analysis (PCA) (Acceleration) | 1 |
| Absolute Value of short-time Fourier Transform (Acceleration) | 1 |
| Power of short-time Fourier Transform (Acceleration) | 1 |
| Power Spectral Centroid (Acceleration) | 1 |

| Features: | Number of Studies: |
|---|---|
| Average of Continuous Wavelet Transform at various Approximation Levels (Acceleration) | 1 |

On the other hand, the distribution of the classification methods used in the studies analyzed is presented in the table 3, verifying that the methods that reports better accuracy are DNN, PNN, kStart, Random Tree, MLP, Logistic Regression, Decision Tree and Random Forest. However, the different types of neural networks are included in the methods with more accuracy, these are DNN, PNN, and MLP, reporting accuracies between 93.86% (MLP) and 98.6% (DNN).

*Table 3 - Distribution of the classification methods used in the studies analyzed*

| Methods: | Number of Studies: | Average of Reported Accuracy: |
|---|---|---|
| Deep Neural Networks (DNN) | 1 | 98.6% |
| Probabilistic Neural Networks (PNN) | 1 | 98% |
| Quadratic classifier (QDA) | 1 | 96.2% |
| kStar | 2 | 96.01% |
| Random Tree | 5 | 94.48% |
| Artificial Neural Networks (ANN) / Multilayer Perceptron (MLP) | 27 | 93.86% |
| Logistic Regression | 7 | 92.18% |
| Decision trees (J48, C4.5) | 30 | 90.89% |
| Random Forest | 12 | 90% |
| Sequential minimal optimization (SMO) | 4 | 89.95% |
| Support Vector Machines (SVM) | 33 | 88.1% |
| ANOVA | 1 | 88% |
| Decision Table | 6 | 87.3% |
| Logistic Model Trees (LMT) | 1 | 85.89% |
| K-Nearest Neighbour (KNN) | 18 | 85.67% |
| Simple Logistic | 1 | 85.05% |
| ZeroR | 1 | 84.8% |
| RBF Network | 1 | 84.4% |
| Hidden Markov model (HMM) | 3 | 84.22% |
| Threshold Based Algorithm (TBA) | 1 | 83% |
| Naïve Bayes | 17 | 82.86% |
| Logit Boost | 1 | 82.39% |
| Least Squares Method (LSM) | 1 | 80% |
| Bayesian network | 4 | 77.81% |
| Linear discriminant analysis (LDA) | 1 | 76.41% |
| J-Rip | 4 | 76.35% |
| IBk | 3 | 76.28% |
| Radial Basis Function Network (RBFNet) | 5 | 72.02% |
| Bagging | 2 | - |
| Sparse Representation Classifier | 2 | - |
| Best-First Tree | 1 | - |
| Rule Based Classifiers | 1 | - |
| Rpart | 1 | - |
| Kernel-Extreme Learning Machine | 1 | - |
| Fuzzy c means | 1 | - |
| Spearman correlation | 1 | - |
| Linear regression | 1 | - |

## 3. Methods

Following the previous research studies analyzed in the section 2, based on the proposed architecture of a framework for the recognition of ADL in [4-6], the methods should be defined for each module of the framework, these are: data acquisition, data processing, data fusion, and artificial intelligence methods. The data processing methods proposed include the data cleaning, data imputation, and feature extraction methods. However, this study assume that the data acquired from the accelerometer is always filled, removing the use of the data imputation methods in this study. Due to the fact that this study only uses the accelerometer available in off-the-shelf mobile devices, the data fusion methods are not needed.

Firstly, the data acquisition method for the architecture implemented is presented in the section 3.1. Secondly, in the presentation of the data processing methods, in section 3.2, the solution is forked in data cleaning (section 3.2.1), and feature extraction methods (section 3.2.2). Finally, the implementation of artificial intelligence methods is detailed in the section 3.3.

### 3.1. Data Acquisition

The data acquisition process was performed with a mobile application developed for Android platform [82, 83] with a BQ Aquarius device [84], where the values of the accelerometer data are received every 10ms. The mobile application acquires 5 seconds of accelerometer data every 5 minutes. The captures where performed with the mobile device in the pocket, and the individuals are aged between 16 and 60 years old with different lifestyles. During the data acquisition, the ADL captured were running, walking, going upstairs, going downstairs, and standing, because these ADL are the most recognized ADLs in the previous studies. The data collected was saved in text files, and it has around 2000 captures for each ADL. The mobile application allows the user to select the ADL performed, for further processing. The data acquired is available in the ALLab MediaWiki [85].

### 3.2. Data Processing

After the data acquisition, the data processing is performed. Firstly, in section 3.2.1, the data cleaning consists in the application of filters to remove the noise captured, and, finally, in section 3.2.2, the features are extracted for the correct identification of the ADL.

#### 3.2.1. Data Cleaning

The first stage of the data processing is named as data cleaning, where the data is filtered, removing the noise. The filter used in the mobile application for the data acquired from accelerometer is the low-pass filter [86]. After the application of the low-pass filter, the noise will be removed, allowing the correct extraction of the features.

#### 3.2.2. Feature Extraction

Based on the filtered data, a set of features was extracted, where the considered features are the 5 greatest distances between the maximum peaks, the Average of the maximum peaks, the Standard Deviation of the maximum peaks, the Variance of the maximum peaks, the Median of the maximum peaks, the Standard Deviation of the raw signal, the Average of the raw signal, the Maximum value of the raw signal, the Minimum value of the raw signal, the Variance of the of the raw signal, and the Median of the raw signal.

### 3.3. Artificial Intelligence

After the extraction of the features, five datasets of features have been created with 2000 records for each ADL, these are:
- **Dataset 1:** Composed by 5 greatest distances between the maximum peaks, Average of the maximum peaks, Standard Deviation of the maximum peaks, Variance of the maximum peaks, Median of the maximum peaks, Standard Deviation of the raw signal, Average of the raw signal, Maximum value of the raw signal, Minimum value of the raw signal, Variance of the of the raw signal, and Median of the raw signal;
- **Dataset 2:** Composed by Average of the maximum peaks, Standard Deviation of the maximum peaks, Variance of the maximum peaks, Median of the maximum peaks, Standard

Deviation of the raw signal, Average of the raw signal, Maximum value of the raw signal, Minimum value of the raw signal, Variance of the of the raw signal, and Median of the raw signal;
- **Dataset 3:** Composed by Standard Deviation of the raw signal, Average of the raw signal, Maximum value of the raw signal, Minimum value of the raw signal, Variance of the of the raw signal, and Median of the raw signal;
- **Dataset 4:** Composed by Standard Deviation of the raw signal, Average of the raw signal, Variance of the of the raw signal, and Median of the raw signal;
- **Dataset 5:** Composed by Standard Deviation of the raw signal, and Average of the raw signal.

Following the previous studies presented in the section 2 and based on the table 3, the method selected for the implementation of the framework for the recognition of ADL is based on the mobile devices' accelerometer is the ANN, because it is one of the most used methods available in the literature, and is has the highest accuracy of the most used methods, and DNN, because is the method available in all studies analyzed with highest accuracy.

After the creation of the datasets, two types of ANN and one type of DNN were applied for the verification of the best type of ANN for the recognition of ADL, these are:
- MLP with Backpropagation, applied with Neuroph framework [13];
- Feedforward Neural Network with Backpropagation, applied with Encog framework [14];
- Deep Neural Networks, applied with DeepLearning4j framework [15].

The MLP with Backpropagation were applied for each dataset in two variants, these are the application with dataset without normalization, and the application with the dataset with the application of a MIN/MAX normalizer [87].

The Feedforward Neural Networks with Backpropagation were applied for each dataset in two variants, these are the application in a dataset without normalization, and the application in a the dataset with the application of a MIN/MAX normalizer [87].

The Deep Neural Networks were applied for each dataset in two variants, these are the application in a dataset with the $L_2$ regularization [88], and the application in a dataset with the $L_2$ regularization and normalized with mean and standard deviation [88, 89].

For the application of these neural networks, three maximum numbers (*i.e.,* 1M, 2M, and 4M) of training iterations have been defined for the identification of the correct number of iterations for the training stage of the neural network.

Finally, the best ANN method should be implemented in the framework for the recognition of ADL using the mobile devices' accelerometer.

## 4. Results

The three types of neural networks, proposed in the previous section, were implemented with the three different frameworks and a training dataset with a total of 10000 records. After the creation of the neural network with Neuroph framework as MLP with Backpropagation, the neural network was tested, and the results obtained are presented in the figure 1. In general, the results obtained with the trained neural networks with 1M, 2M, and 4M iterations, and different sets of features have very low accuracy (between 20% and 40%) with data without normalization, presented in figure 1-a, and very low accuracy (between 20% and 30%) with normalized data, presented in figure 1-b.

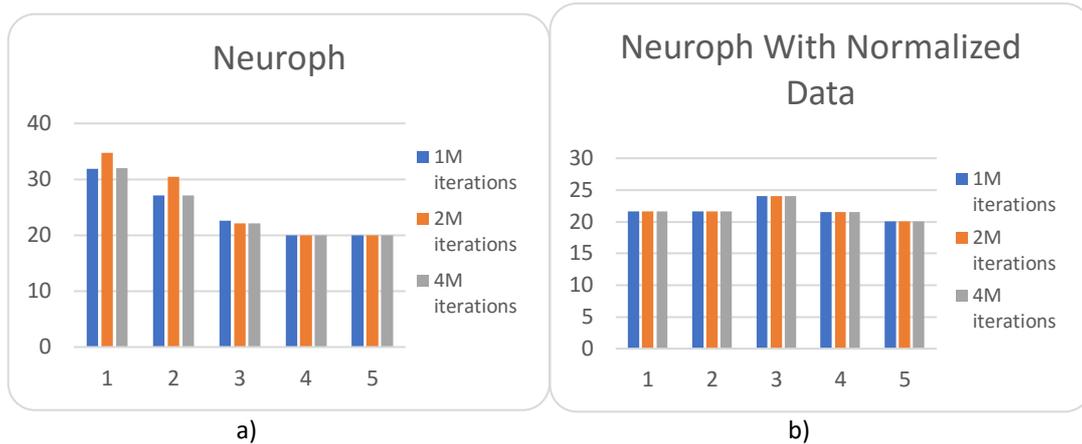

*Figure 1 –Results obtained with Neuroph framework for the different datasets (horizontal axis) and different maximum number of iterations (series), obtaining the accuracy in percentage (vertical axis). The figure a) shows the results with data without normalization. The figure b) shows the results with normalized data.*

After the test of the MLP with Backpropagation, the Feedforward Neural Network with Backpropagation was created with Encog framework, obtaining the results presented in the figure 2. In general, the results obtained with the trained neural networks with 1M, 2M, and 4M iterations, and different sets of features have very low accuracy (between 20% and 40%) with data without normalization, presented in figure 2-a, where, as exceptions, the neural networks trained with the dataset 3 with 2M iterations obtains an accuracy around 50%, and with the dataset 5 with 4M iterations obtains an accuracy around 75%. On the other hand, when the data is normalized, the results presented in figure 2-b, shows that the reduction of the number of the features in the datasets increases the accuracy of the neural network.

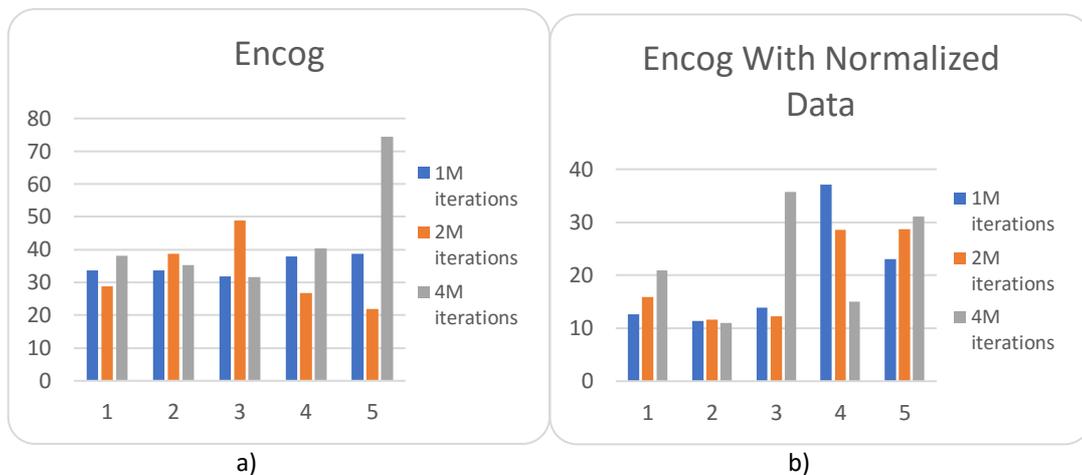

*Figure 2 –Results obtained with Encog framework for the different datasets (horizontal axis) and different maximum number of iterations (series), obtaining the accuracy in percentage (vertical axis). The figure a) shows the results with data without normalization. The figure b) shows the results with normalized data.*

After the verification that results obtained with MLP with Backpropagation, and Feedforward Neural Network are not satisfactory, the Deep Neural Network was created with DeepLearning4j framework, obtaining the results presented in the figure 3. In general, the results obtained with the trained neural networks with 1M, 2M, and 4M iterations, and the different sets of features have an accuracy higher than 70%, but, with data without normalization (figure 3-a), the results obtained with the datasets 1 and 2 are above the expectations with an accuracy lower than 40%, and, with the normalized data (figure 3-b), the results obtained are higher with dataset 1, decreasing with the reduction of the number of features in the dataset.

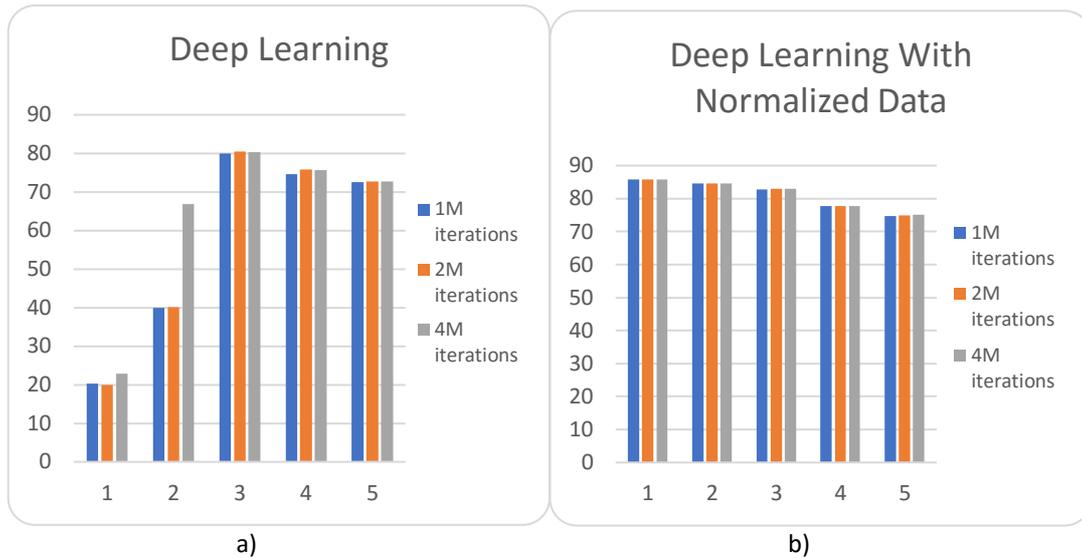

*Figure 3 –Results obtained with DeepLearning4j framework for the different datasets (horizontal axis) and different maximum number of iterations (series), obtaining the accuracy in percentage (vertical axis). The figure a) shows the results with data without normalization. The figure b) shows the results with normalized data.*

The maximum accuracies obtained with the MLP with Backpropagation, Feedforward Neural Networks with Backpropagation, and Deep Neural Networks are shown in the table 4, concluding that the results obtained by MLP with Backpropagation and Feedforward Neural Networks with Backpropagation are not satisfactory, obtaining best accuracies with Deep Neural Networks.

*Table 4 - Best accuracies obtained with the different frameworks, datasets and number of iterations*

|  | FRAMEWORK | DATASET | ITERATIONS NEEDED FOR TRAINING | BEST ACCURACY ACHIEVED (%) |
|---|---|---|---|---|
| NOT NORMALIZED DATA | NEUROPH | 1 | 2M | 34.76 |
|  | ENCOG | 5 | 4M | 74.45 |
|  | DEEP LEARNING | 3 | 4M | 80.35 |
| NORMALIZED DATA | NEUROPH | 3 | 1M | 24.03 |
|  | ENCOG | 4 | 1M | 37.07 |
|  | DEEP LEARNING | 1 | 4M | 85.89 |

In conclusion, the type of neural networks that should be used in the framework for the identification of ADL is the Deep Neural Network (Deep Learning), because the results are constantly higher than the others, and the accuracy of the recognition reports reliable results in the tests performed.

## 5. Discussion and Conclusions

This paper presents a method for the identification of several ADL, including running, walking, going upstairs, going downstairs, and standing, comparing the results obtained with different types of neural networks. The development of the method presented in this paper was based in [4-6], including only the data acquisition, data processing with data cleaning and feature extraction, and artificial intelligence methods, requiring low processing for the correct implementation in the mobile devices.

The comparison between MLP with Backpropagation, applied with Neuroph framework [13], Feedforward Neural Network with Backpropagation, applied with Encog framework [14], and Deep Neural Networks, applied with DeepLearning4j framework [15], reports that the use of Deep Neural Networks increases the accuracy of the recognition of the ADL. The datasets used in the neural networks were composed by 10000 records, *i.e.*, 2000 records for each ADL. The best results are obtained with Deep Neural Networks with $L_2$ regularization and normalized data.

The low accuracies verified with MLP with Backpropagation, and Feedforward Neural Network with Backpropagation are related to the fact of the neural networks created are overfitting, and the possible solutions are the acquisition of more data, the stopping of the training when the network error increases for several iterations, the application of dropout regularization, the application of $L_2$ regularization, the application of the batch normalization, or the reduction of the number of features in the neural network.

The number of the maximum iterations may influence the training of the neural network, and, in some cases, it also increases the accuracy of the neural network, but the influence if the number of iterations are not substantial.

In conclusion, the method implemented in the framework for the recognition of the ADL using only the accelerometer sensor available in off-the-shelf mobile devices should be based in Deep Neural Networks, applied with DeepLearning4j framework [15], because it achieves an accuracy above 80% with a neural network trained with all features proposed in this study, these are the 5 greatest distances between the maximum peaks, the Average of the maximum peaks, the Standard Deviation of the maximum peaks, the Variance of the maximum peaks, the Median of the maximum peaks, the Standard Deviation of the raw signal, the Average of the raw signal, the Maximum value of the raw signal, the Minimum value of the raw signal, the Variance of the of the raw signal, and the Median of the raw signal. This research proves the reliability of the use of ANN for the identification of the ADL using the accelerometer.

## Acknowledgements

This work was supported by FCT project **UID/EEA/50008/2013** (*Este trabalho foi suportado pelo projecto FCT UID/EEA/50008/2013*).

The authors would also like to acknowledge the contribution of the COST Action IC1303 – AAPELE – Architectures, Algorithms and Protocols for Enhanced Living Environments.